  \documentclass[fp]{jpsj3}
\usepackage{txfonts}
\usepackage{color}
\newcommand{\bp}{\boldsymbol{p}}
\newcommand{\bi}{\boldsymbol{i}}
\newcommand{\bj}{\boldsymbol{j}}
\newcommand{\ve}{\varepsilon}
\newcommand{\be}{\begin{equation}}
\newcommand{\ee}{\end{equation}}
\newcommand{\del}{\partial}
\newcommand{\uya}{\uparrow}
\newcommand{\dya}{\downarrow}
\newcommand{\udya}{\uparrow\downarrow}

\title{Analysis of Particle Transfer by Periodic Lattice Modulation
for Ultracold Fermionic Atom Systems in Three Dimensional Optical Lattice
}

\author{\name{Motoyoshi \surname{Inoue}}\thanks{E-mail: local-fortokyo@suou.waseda.jp}, \name{Yusuke \surname{Nakamura}}, and \name{Yoshiya \surname{Yamanaka}}}
\inst{Department of Electronic and Photonic Systems, Waseda University 3-4-1 Okubo Shinjuku-ku Tokyo 169-8555, Japan} 

\abst{
We analyze a ultracold fermionic atom system in a three dimensional optical lattice with a confinement harmonic potential, using the Hubbard model, and time-dependent
 Gutzwiller variational approach for numerical calculation.
Our study is focused on the time evolution of the particle transfer when the lattice
 potential is modulated by adding a periodic one. The choice of the parameters such as the 
modulation frequency and amplitude and the particle number affects the particle transfer.
We calculate the time evolution of the variance in the particle distribution, and show
its dependence on the parameters. The lattice modulation turns out to work effectively
 in order to control the particle transfer, and will be a useful method in experiments
for fermionic atom systems.}

\kword{optical lattice modulation, particle transfer dynamics, three dimensional inhomogeneous system, Fermi Hubbard model, Gutzwiller variational approach, insulator transition}

\begin{document}
\maketitle

\section{Introduction}

The ultracold fermionic atom system in an optical lattice attracts attention, since
some intriguing phenomena such as Mott insulator and anti-ferromagnetism have been 
observed in the experiments and its analyses will bring hints to understand
many problems in solid state physics.
The systems, both bosonic and fermionic, are well described by the Hubbard model, 
whose numerical analysis are performed in various methods, i.e., Gutzwiller variational approach (GVA)\cite{GW1,GW2,GW3}, dynamical mean-field theory (DMFT)\cite{D-mean-field1,D-mean-field2,D-mean-field3}, density matrix renormalization group (DMRG) method\cite{fermionfirst,DMRG1,DMRG2,DMRG3}, quantum Monte Carlo (QMC) method \cite{fermidoublonlinear,QMC1,QMC2} and so on.

To control the particle distribution of fermionic atoms in an optical lattice is
not easy, since the transfer of fermionic atoms in an optical lattice is slower than that of bosonic one due to the Pauli blocking.  The experiment to realize the Mott insulator transition 
in fermionic case is more difficult than those of bosonic ones in general. 
Actually it takes several tens of milliseconds to configure the ground state of Mott insulator in the bosonic experiment \cite{bosonMottexp}, but several hundreds of milliseconds in the fermionic one \cite{fermiexp}.

As a promising way to manage to control the particle distribution of fermionic atoms in 
an optical lattice, the periodic lattice modulation, which means an addition of 
an oscillating lattice to the constant one, has been studied in the theoretical analysis for fermionic atom system using
linear response \cite{fermionfirst,fermidoublonlinear}
 and in the experiment \cite{fermi-doublonexp}.
 When the lattice modulation is 
under consideration, there are a few parameters, i.e., the modulation amplitude and frequency
which can be varied easily in experiments and are expected to affect the transfer crucially.
The previous studies show that the particle transfer is stimulated by the periodic lattice modulation, especially, for the resonant frequency of the lattice modulation and for the large modulation amplitude, though they are limited to the case
 close to the half-filling 
and to the homogeneous system without a confinement harmonic potential. 
The responses to the lattice modulation will be different for the homogeneous and inhomogeneous systems. The purpose of this paper is to analyze the response or particle transfer of fermionic
atom system in a three dimensional optical lattice modulated by a periodic potential with a 
confinement harmonic potential.
Since the experimental parameters are well controllable in ultracold atom systems, 
the analysis of the lattice modulation response in the inhomogeneous system will enable us to 
understand some new physical properties of complicated fermionic systems.

 We analyze our three dimensional system, using the Hubbard model. 
The DMRG method for higher dimensional system remains to be established.
The numerical calculations for the inhomogeneous three dimensional system are
computationally expensive for the QMC or DMFT, so they are not realistic choices. 
 The cost of numerical calculation in the GVA is lower than those in the other methods
and reasonable, so GVA is the most suitable for our calculation.
 We numerically calculate the dynamical evolution under the time-dependent GVA \cite{TD-GW1,TD-GW2}. 

 We investigate the time evolution of the particle distribution, explicitly
that of the variance in the particle distribution as an indicator of
particle transfer, 
starting from some spatially inhomogeneous distributions
and under the periodic lattice modulation
The results are shown for various values of the parameters, the lattice modulation amplitude 
and frequency, the initial phase of the modulation and the total particle number.
Finally we argue the choices of the parameter values to promote the particle transfer to the outer
sites, and this knowledge will be helpful in designing experiments of fermionic systems.

 This paper is structured as follows.
 In Sec.~2, we develop our theoretical approach using the Fermi Hubbard model and the Gutzwiller 
approximation. The section is separated into two parts, the Gutzwiller wave function (GWF) of the 
ground states and time-dependent GVA. The former is used for the initial condition in the latter, 
and the latter yields the time evolution of the system.
The results of numerical calculations are presented in Sec.~3.
Section 4 is devoted to summary and conclusion.

\section{Gutzwiller variational approach}

\subsection{Ground state}
In this subsection, we briefly review the Gutzwiller variational approach (GVA)\cite{GW1,GW2,GW3} to obtain
the ground state of the Hubbard model for three dimensional system. 
The Hubbard Hamiltonian is
\be
\hat{H}=-J\sum_{\langle \bi , \bj \rangle,s}\hat{a}_{\bi,s}^\dagger\hat{a}_{\bj,s}+\sum_{\bi,s}
(\nu_{\bi}-\mu){\hat n}_{\bi,s}+U\sum_{\bi}\hat{n}_{\bi,\uya}\hat{n}_{\bi,\dya}
\,,
\qquad {\hat n}_{\bi,s}=
\hat{a}_{\bi,s}^\dagger\hat{a}_{\bi,s}
\,.\label{H_E}
\ee
The first term is the hopping one, and there 
the hopping coefficient $J$ is independent both of the site 
indices $\bi=(i_x, i_y, i_z)$ and the spin index $s = \uya,\,\dya$. The sum index $\langle \bi, \bj\rangle$ 
denotes the sum over the nearest neighbor sites. The second term represents the confinement potential and the chemical one, and the third one does the on-site interaction.

In GVA, the variational trial function, called the Gutzwiller wave function (GWF), is taken as
\be
|\Psi_G\rangle=\prod_{\bi}\hat{P}_{\bi}|\Psi_0\rangle=\prod_{\bi}\sum_\Gamma\lambda_{\bi,\Gamma}\hat{m}_{\bi,\Gamma}|\Psi_0\rangle\,,
\ee
where $\hat{m}_{\bi,\Gamma}=|\bi,\Gamma\rangle\langle \bi,\Gamma|$ is the projection operator of particle state in site $\bi$, 
 and $\Gamma$ stands for one of the four single site states \{ $0, \uya, \dya, \uya\dya$ \}. 
The key point of GVA is that there are two kinds of variational quantities, {\it i.e.}\,,
the parameter $\lambda_{\bi,\Gamma}$ and the wave function $|\Psi_0\rangle$. 
The former is chosen to minimize the total energy, while the latter is so to minimize the kinetic energy, as will be seen below. 
The expectation value of any product of operators, which is boson-like and local in site,
 with respect to
$|\Psi_0\rangle$ is approximated (Gutzwiller approximation) as
\be \label{GWapproximation}
	\langle \Psi_0| \hat A_{\bi} \hat B_{\bj} |\Psi_0\rangle = 
	\langle \Psi_0| \hat A_{\bi} |\Psi_0\rangle \langle \Psi_0|  \hat B_{\bj} |\Psi_0\rangle \,,
\ee
when $\bi\ne \bj$.
It implies
\be
	\langle \Psi_G|\Psi_G\rangle = 
	\prod_{\bi}\langle\Psi_0| \hat{P}_{\bi}^\dagger \hat{P}_{\bi} |\Psi_0\rangle =
	\prod_{\bi} \left(\sum_{\Gamma}m_{\bi,\Gamma}\right)\,,
\ee
with 
\be	\label{definition_of_m}
	m_{\bi,\Gamma}=|\lambda_{\bi,\Gamma}|^2\langle\Psi_0|\hat{m}_{\bi,\Gamma}|\Psi_0\rangle \,.
\ee
For the normalization of $|\Psi_G\rangle$, we set
\be
	\sum_{\Gamma}m_{\bi,\Gamma} = 1 \,.
\ee

It is also assumed that $|\Psi_0\rangle$ is not spin entangled, 
\be 
|\Psi_0\rangle=|\psi_\uya\rangle \otimes |\psi_\dya\rangle \, .
\label{notentangled}
\ee 
Then from the Gutzwiller approximation (\ref{GWapproximation}) and the identity
\be
\hat{P}^\dagger_{\bi}\hat{a}_{\bi,s}\hat{P}_{\bi}
=\left\{ \lambda_{\bi,s}\lambda^\ast_{\bi,0}(1-{\hat n}
_{\bi,\bar{s}})+\lambda_{\bi,\udya}\lambda^\ast_{\bi,\bar{s}}
{\hat n}_{\bi,\bar{s}}\right\}\hat{a}_{\bi,s}\,,
\ee
where ${\bar s}$ denotes the opposite spin to $s$,
we obtain the expectation value of the total energy,
\be
	\langle\Psi_G|\hat{H}|\Psi_G\rangle = 
	-J\sum_{\langle \bi ,\bj\rangle, s}Z_{\bi,s}^*Z_{\bj,s}
	\langle \Psi_0|\hat{a}_{\bi,s}^\dagger \hat{a}_{\bj,s}|\Psi_0\rangle
	+\sum_{\bi}\left\{\left(\nu_{\bi}-\mu\right)\left(m_{\bi,\uya}+m_{\bi,\dya}+2m_{\bi,\udya}\right)+
	Um_{\bi,\udya}\right\}\,,\label{allH-expectation}
\ee
with 
\be
Z_{\bi,s}=\lambda_{\bi,s}\lambda^\ast_{\bi,0}(1-n_{\bi,\bar{s}}^0)+\lambda_{\bi,\udya}\lambda^\ast_{\bi,\bar{s}}n_{\bi,\bar{s}}^0\,,
	\qquad
n_{\bi,s}^0=\langle\Psi_0|\hat{n}_{\bi,s}|\Psi_0\rangle\,.
\ee
This specific form of the total energy makes it possible 
to fix the variational parameter $\lambda_{\bi,s}$ to be real without loss of generality.
For convenience we take the real parameter $m_{\bi,\Gamma}$ as the variational parameter instead
of $\lambda_{\bi,s}$, both are related to each other by Eq.~(\ref{definition_of_m}).
The quantity $Z_{\bi,s}$ is rewritten as
\be
	\label{Z}
	Z_{\bi,s} = 
	\sqrt{ \frac{m_{\bi,0} \; m_{\bi,s}}{n_{\bi,s}^0(1-n_{\bi,s}^0)}}+
	\sqrt{ \frac{m_{\bi,{\bar s}}\;m_{\bi,\uya\dya}}{n_{\bi,s}^0(1-n_{\bi,s}^0)}} \,.
\ee

The variational wave function $|\Psi_0\rangle$ is determined by the requirement that it 
should minimize the kinetic energy,
\be
	\langle\Psi_G|\hat{H}_J|\Psi_G\rangle = 
	-J\sum_{\langle \bi ,\bj\rangle, s}Z_{\bi,s}Z_{\bj,s}
	\langle \Psi_0|\hat{a}_{\bi,s}^\dagger \hat{a}_{\bj,s}|\Psi_0\rangle \,.
\ee
This is accomplished by diagonalizing the effective hopping Hamiltonian
\be
\hat{H}_{\rm J}=\sum_s\hat{H}_{{\rm J},s}=-J\sum_{\langle \bi,\bj\rangle, s}Z_{\bi,s}Z_{\bj,s}\hat{a}_{\bi,s}^\dagger\hat{a}_
{\bj,s}\,.\label{H_Jda}
\ee
Because there is no cross-term of up- and down-spins in Eq.~(\ref{H_Jda}), 
$|\Psi_0\rangle$ is given by the direct product of the spin states, which is consistent 
with the assumption Eq.~(\ref{notentangled}).
The coefficient hopping Hamiltonian $\hat{H}_{{\rm J},s}$ is diagonalized by the
unitary transformation $\alpha_{\bp,s} =\sum_{\bi} U_{\bp, \bi,s} a_{\bi,s}$ as
\be
	\hat{H}_{{\rm J},s} = 
	\sum_{\bp}\ve_{\bp,s}\hat{\alpha}_{\bp,s}^\dagger\hat{\alpha}_{\bp,s} \,,
\ee
and the ground state $|\Psi_0\rangle$ is determined as
\be
	|\Psi_0\rangle = \left( \prod_{\bp}^{\ve_{\bp,\uya} < 0} \alpha_{\bp,\uya}^\dagger\right) |0\rangle_\uya 
	\otimes \left( \prod_{\bp}^{\ve_{\bp,\dya} < 0} \alpha_{\bp,\dya}^\dagger \right)|0\rangle_\dya  \,,
\ee
where $|0\rangle_s$ is the vacuum of ${\alpha}_{\bp,s}$.
Thus we obtain 
\be
	\langle \Psi_0|\hat{a}_{\bi,s}^\dagger \hat{a}_{\bj,s}|\Psi_0\rangle = 
	\sum_s \sum_{\bp}^{\ve_{\bp,s} < 0} U_{\bp, \bi,s} U_{\bp, \bj,s} \,.
\ee

The quantity $Z_{\bi,s}$ and consequently the unitary matrix $U_{\bp, \bi,s}$
are functions of $m_{\bi,\Gamma}$, which in turn depends on $|\Psi_0\rangle$ as in 
Eq.~(\ref{definition_of_m}).  So the variational quantities
$m_{\bi,\Gamma}$ and $|\Psi_0\rangle$ have to be determined in a self-consistent manner.
\subsection{Dynamics}
We next review the time-dependent GVA \cite{TD-GW1,TD-GW2}. 
To describe the dynamics of the present system, the stationary GWF is extended straightforwardly to the time-dependent one 
by making the variational parameter $\lambda_{\bi,\Gamma}$ complex and time-dependent,
\be
	|\Psi_G(t)\rangle=\prod_{\bi}\sum_{\Gamma}\lambda_{\bi,\Gamma}(t)\hat{m}_{\bi,\Gamma}|\Psi_0\rangle\,.\label{TDGWF}
\ee
The equation of time evolution is derived by the variational principle for the action,
\be
S=\int\! dt\,
\langle\Psi_G(t)|i\hbar\frac{d }{d t}-\hat{H}|\Psi_G(t)\rangle\label{sayou}\,.
\ee
The integrand  is manipulated as 
\begin{align}
	\langle\Psi_G(t)|i\hbar\frac{d }{d t}-\hat{H}|\Psi_G(t)\rangle
	= \hbar \sum_{\bi,\Gamma}\biggl( \prod_{\bj\ne\bi} \sum_{\Gamma'} m_{\bj,\Gamma'}(t)\biggr) \left(\dot{\phi}_{\bi,\Gamma}(t) m_{\bi,\Gamma}+\frac{i}{2}\dot{m}_{\bi,\Gamma}(t)\right)-E[m_{\bi,\Gamma},\phi_{\bi,\Gamma}] \,,
\end{align}
with $\phi_{\bi,\Gamma}(t) = -\arg \lambda_{\bi,\Gamma}(t)$, $	m_{\bi,\Gamma}(t) = |\lambda_{\bi,\Gamma}(t)|^2\langle\Psi_0|\hat{m}_{\bi,\Gamma}|\Psi_0\rangle$, and $E[m_{\bi,\Gamma}, \phi_{\bi, \Gamma},t ]  =\langle\Psi_G(t)|\hat{H}|\Psi_G(t)\rangle \,.$
For the conservation of the normalization $\langle\Psi_G(t)|\Psi_G(t)\rangle=1$, we require
\be	\label{requirement}
	\sum_{\Gamma} m_{\bi,\Gamma}(t) = 1\,,
\ee
whose consistency will be confirmed later.
Then the action is simplified as 
\be
	S=\int\!dt\,  \left(\sum_{\bi,\Gamma}
 \hbar \dot\phi_{\bi,\Gamma}(t) m_{\bi,\Gamma}(t) - E[m_{\bi,\Gamma}, \phi_{\bi, \Gamma}]\right) \,,
\ee
which implies that $\phi_{\bi,\Gamma}$ and $m_{\bi,\Gamma}$ are a pair of canonical variables.
Their equations of motion are given by
\be	\label{EoM}
	\hbar \dot{m}_{\bi,\Gamma}		= -\frac{\del E}{\del \phi_{\bi,\Gamma}}\,,\qquad
	\hbar \dot{\phi}_{\bi,\Gamma}	= \frac{\del E}{\del m_{\bi,\Gamma}}\,.
\ee

The total energy is expressed in the same form as Eq.~(\ref{allH-expectation}) but the coefficients become complex and time-dependent:
\begin{align}
	E[m_{\bi,\Gamma}, \phi_{\bi, \Gamma}] &= 
	-J\sum_{\langle \bi ,\bj\rangle, s}Z_{\bi,s}^*(t)Z_{\bj,s}(t)
	\langle \Psi_0|\hat{a}_{\bi,s}^\dagger \hat{a}_{\bj,s}|\Psi_0\rangle \nonumber \\
&\quad 	+\sum_{\bi}\left\{\left(\nu_{\bi}-\mu\right)\left(m_{\bi,\uya}(t)+m_{\bi,\dya}(t)+2m_{\bi,\udya}(t)\right)+
	Um_{\bi,\udya}(t)\right\}\,.\label{T-expectation}
\end{align}
with
\be
	Z_{\bi,s}(t)=\sqrt{\frac{m_{\bi,0}(t)m_{\bi,s}(t)}{n_{\bi,s}^0(1-n_{\bi,s}^0)}}
	e^{i(\phi_{\bi,0}(t)-\phi_{\bi,s}(t))}
	+\sqrt{\frac{m_{\bi,\bar{s}}(t)m_{\bi,\udya}(t)}{n_{\bi,s}^0(1-n_{\bi,s}^0)}}
	e^{i(\phi_{\bi,\bar{s}}(t)-\phi_{\bi,\udya}(t))}\,.
\ee
Because of the relations
\be
	\frac{\del Z_{\bi,s}}{\del \phi_{\bi,0}}+\frac{\del Z_{\bi,s}}{\del \phi_{\bi,s}}=0\,,\qquad
	\frac{\del Z_{\bi,s}}{\del \phi_{\bi,\udya}}+\frac{\del Z_{\bi,s}}{\del \phi_{\bi,\bar{s}}}=0\,,
\ee
we obtain 
\be
	\sum_{\Gamma}\frac{\del E}{\del \phi_{\bi,\Gamma}}=0\,.
\ee
This, with Eq.~(\ref{EoM}), leads to Eq.~(\ref{requirement}), and the consistency of 
Eq.~(\ref{requirement}) has been checked.
We can also confirm the energy conservation
\be
	\frac{d}{dt}E[m_{\bi,\Gamma},\phi_{\bi,\Gamma}] = 
	\sum_{\bi,\Gamma} \left(\frac{\del E}{\del m_{\bi,\Gamma}}\dot{m}_{\bi,\Gamma}+
	\frac{\del E}{\del \phi_{\bi,\Gamma}}\dot{\phi}_{\bi,\Gamma} \right)=0\,.
\ee

\section{Numerical Results}

In this section, we present the results of numerical calculations
according to the variational formulations in the previous section. Our main interest is
in the dynamics of particle transfer, and we devote ourselves below
to study it under the periodic lattice modulation, 
meaning that an oscillating sinusoidal  lattice 
potential is added for $t>0$ to the constant one.  Equation~(\ref{EoM}) is solved numerically,
and it will be shown how the particles are transferred for various values of the amplitude 
and frequency of the lattice modulation.

\subsection{Parameters}

In our numerical calculations below, the spatial site number is $18^3$. 
The confinement harmonic potential is assumed to be isotropic 
with its center at that of the lattice system,
 $\bi_0=\frac{18-1}{2}(1,1,1)=(8.5,8.5,8.5)$, and to have the fixed strength,
\be
\nu_{\bi}=
0.2 E_{\rm R}\,\left|\bi-\bi_0 \right|^2 \,,
\ee
where $E_{\rm R}=h^2/2m\lambda^2$ is the recoil energy and
 $\lambda$ is the wavelength of the laser light, corresponding to a lattice period $a=\lambda/2$.
The constant lattice amplitude before the lattice potential is modulated
is put as ${\bar V}_0=6.5E_{\rm R}$, 
and 
this value is not so large that the stable insulator could be formed.
 The s-wave scattering length is $a_s=240a_0$, where $a_0$ is Bohr radius. We quote these parameters from the experimental ones \cite{fermiexp}.
%

The lattice is modulated as a sudden switch-on of the sinusoidal potential at $t=0$:
\be
V_0(t)=
\begin{cases}
{\bar V}_0 & t<0 \,,\\
{\bar V}_0-\delta V \sin \left[\omega t+\theta\right] & t>0 \, , 
\end{cases}
\label{lattice}
\ee
where $\delta V$ and $\omega$ are referred to the modulation amplitude and frequency, respectively.
It is expected that the frequency resonant with the interaction energy invokes
 the particle transport most effectively. 
Note that the parameters of  the Hubbard model, $J(t)$ and $U(t)$, are time-dependent as
 follows \cite{Hubbard-para}, since
they are functions of $V_0(t)$,
\begin{align}
U(t)&=4\sqrt{2\pi}\left(a_s/\lambda\right)\left(\tilde{v}(t)\right)^{3/4}E_{\rm R}\,,\label{paraU}
\\
J(t)&=4/\sqrt{\pi}\left (\tilde{v}(t)\right ) ^{3/4}\exp\left[-2\left(\tilde{v}(t)\right)^{1/2}\right ]E_{\rm R}\,,\label{paraJ}
\end{align}
with $\tilde{v}(t)=V_0(t)/E_{\rm R}$.
Though $U(t)$ is time-dependent [see Eq.(\ref{paraU})], we define the resonant frequency
$\omega_{\rm re}= {\bar U}/\hbar$, using the mean interaction strength ${\bar U}= U(t)$ for $t<0$. 
The phase in Eq.~(\ref{lattice}) is taken either $\theta=0$ or $\pi$.
 In this paper, we calculate for various strengths of the modulation amplitude, 
$\delta V/{\bar V}_0=0.1, 0,2, 0.3, 0.4$, and 0.5, and for various modulation frequencies 
$\omega/\omega_{\rm re}=0.5, 0.8, 1.0, 1.5, 2.0$, and $2.5$.
\begin{figure}
\includegraphics[width=7.2cm]{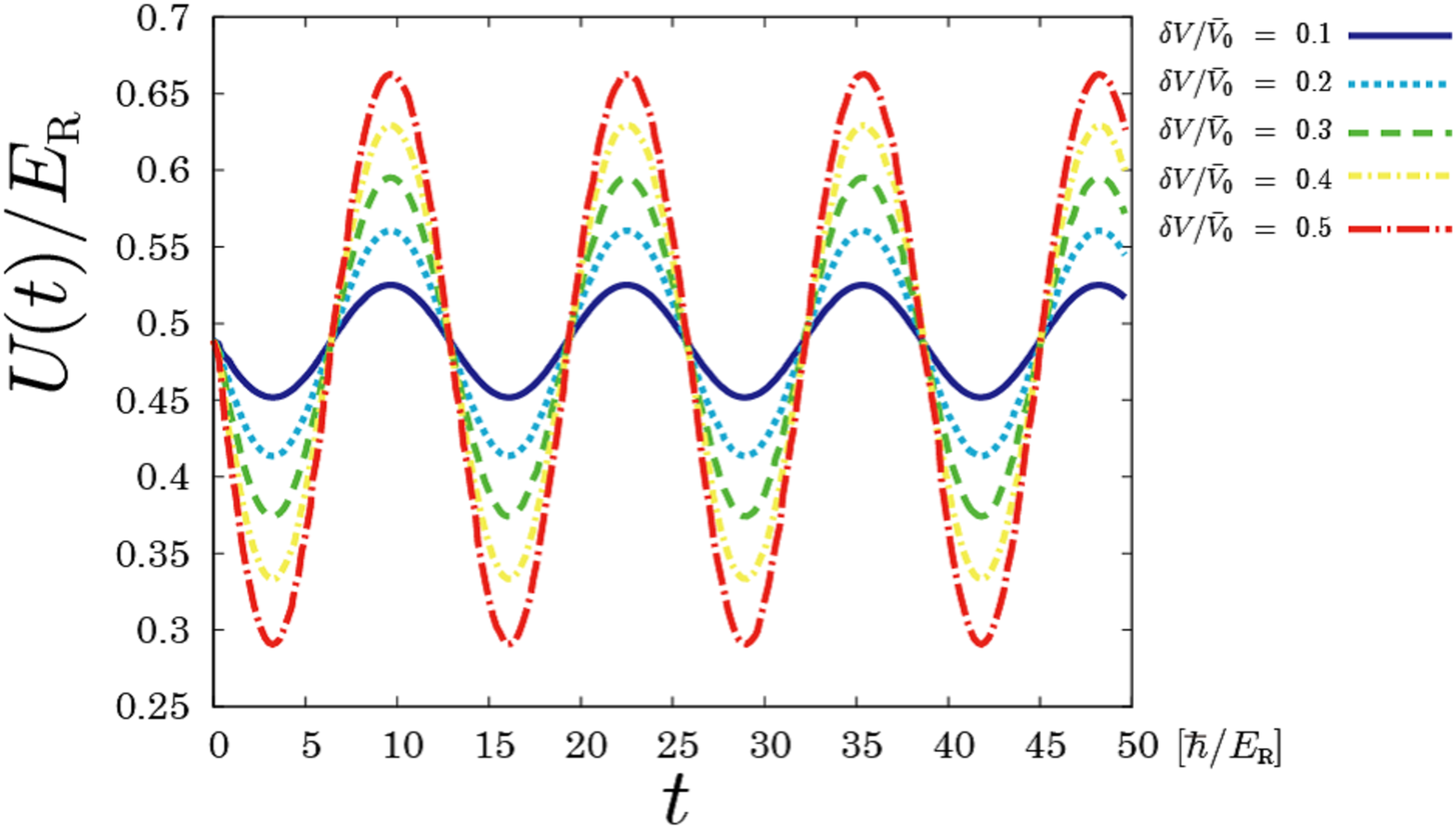}\qquad\includegraphics[width=7.2cm]{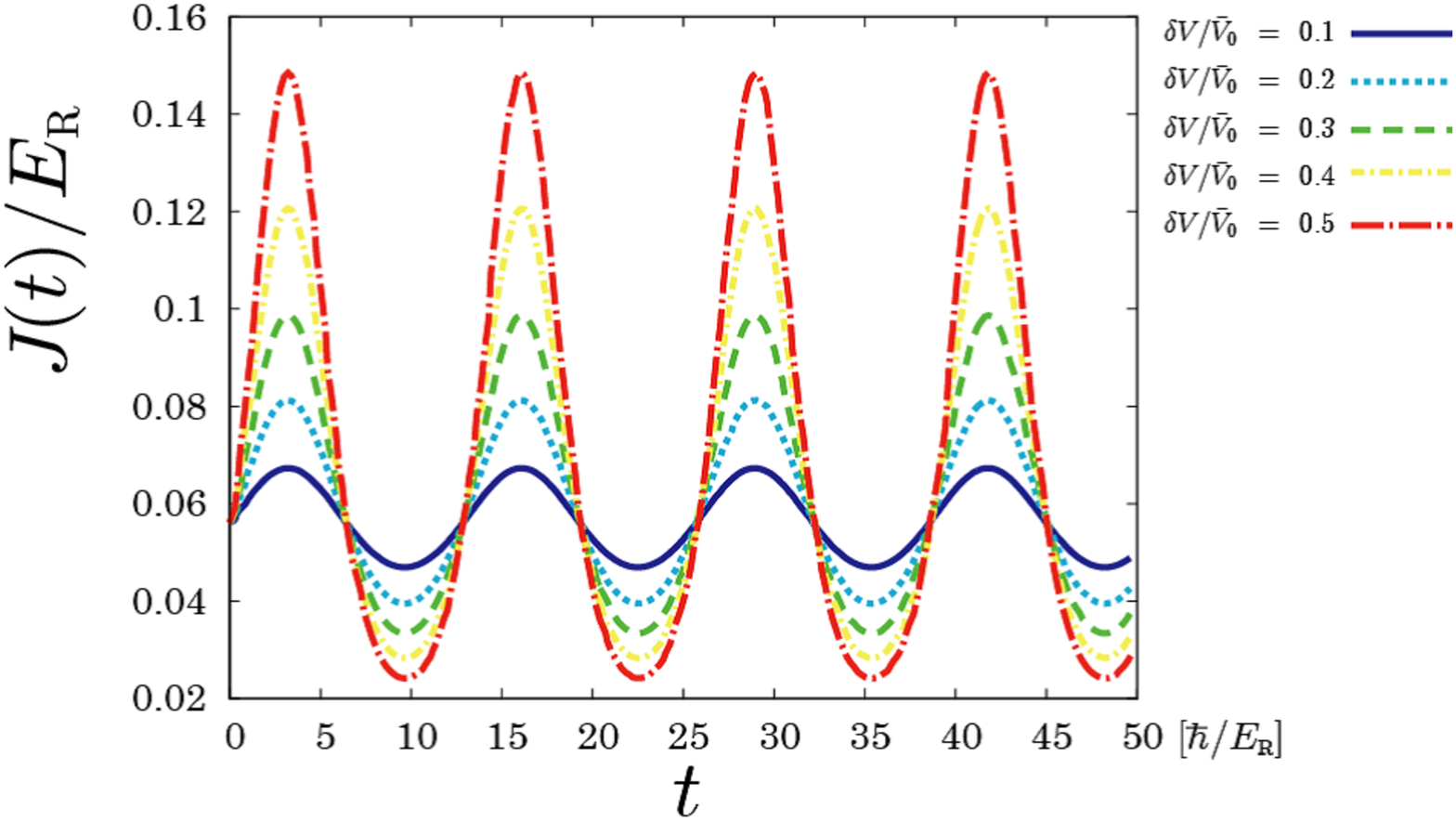}
\caption{Temporal behaviors of $U(t)$ and $J(t)$ for the modulation amplitude
 $\delta V/{\bar V_0}=0, 0.1, 0,2, 0.3, 0.4$, and $ 0.5$. The period of the oscillation is 
12.9$\hbar/E_{\rm R}$.}\label{J-Uchange}
\end{figure}

The temporal behaviors of $U(t)$ and $J(t)$ in Eqs.~(\ref{paraU}) and (\ref{paraJ})
are plotted in Fig. \ref{J-Uchange}.
The hopping coefficient $J(t)$ behaves more intricately than the interaction one $U(t)$.
This is due to the presence of the factor $\exp\left[-2\left(\tilde{v}(t)\right)^{1/2}\right ]$
in the former. We remark that the local maximum value of $J(t)$
is larger for larger $\delta V$, though it behaves almost sinusoidally for small $\delta V$,
reflecting $ (\tilde{v}(t)) ^{3/4} $.

\begin{figure}
\includegraphics[width=7.2cm]{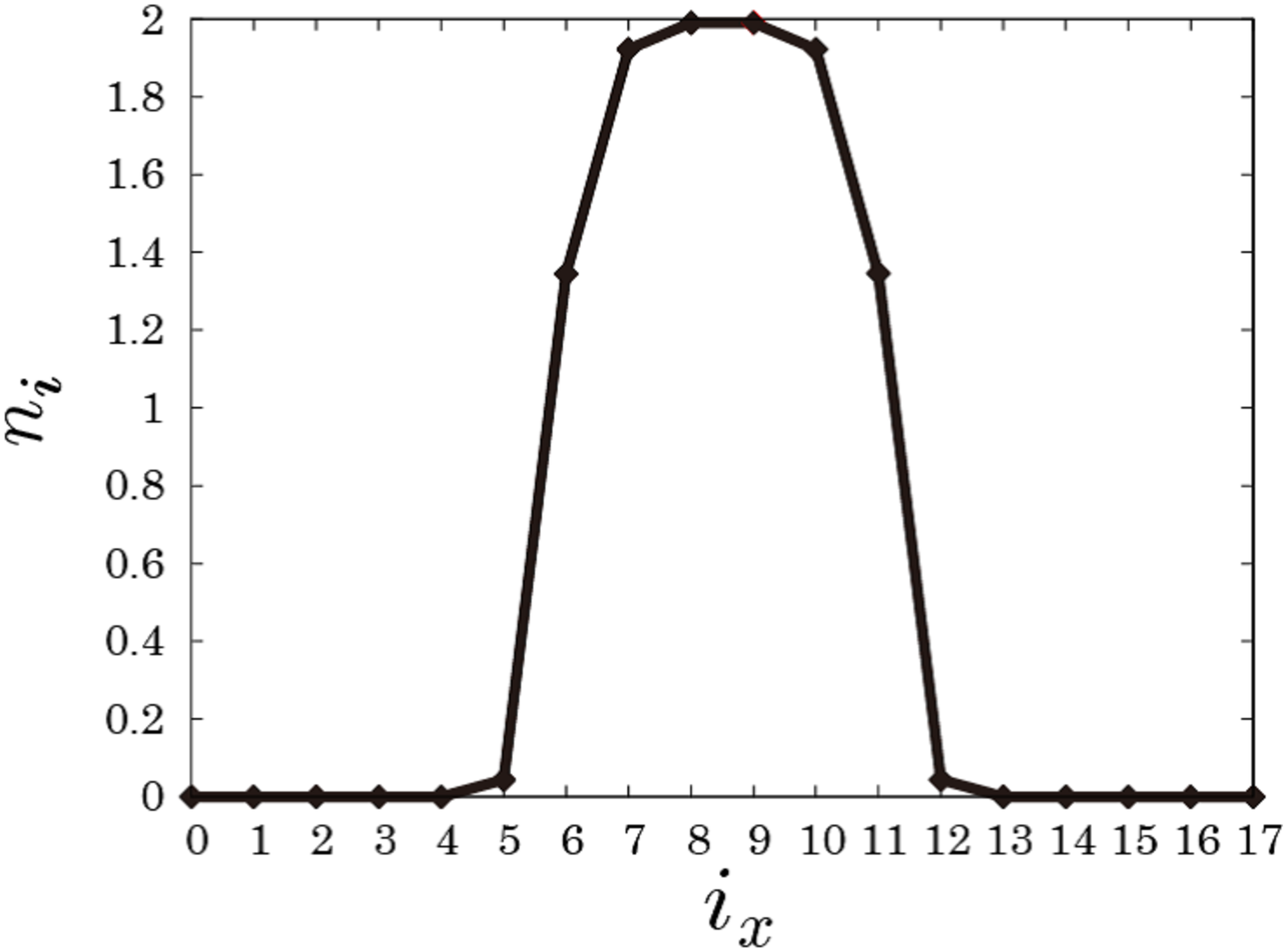}\, \includegraphics[width=7.2cm]{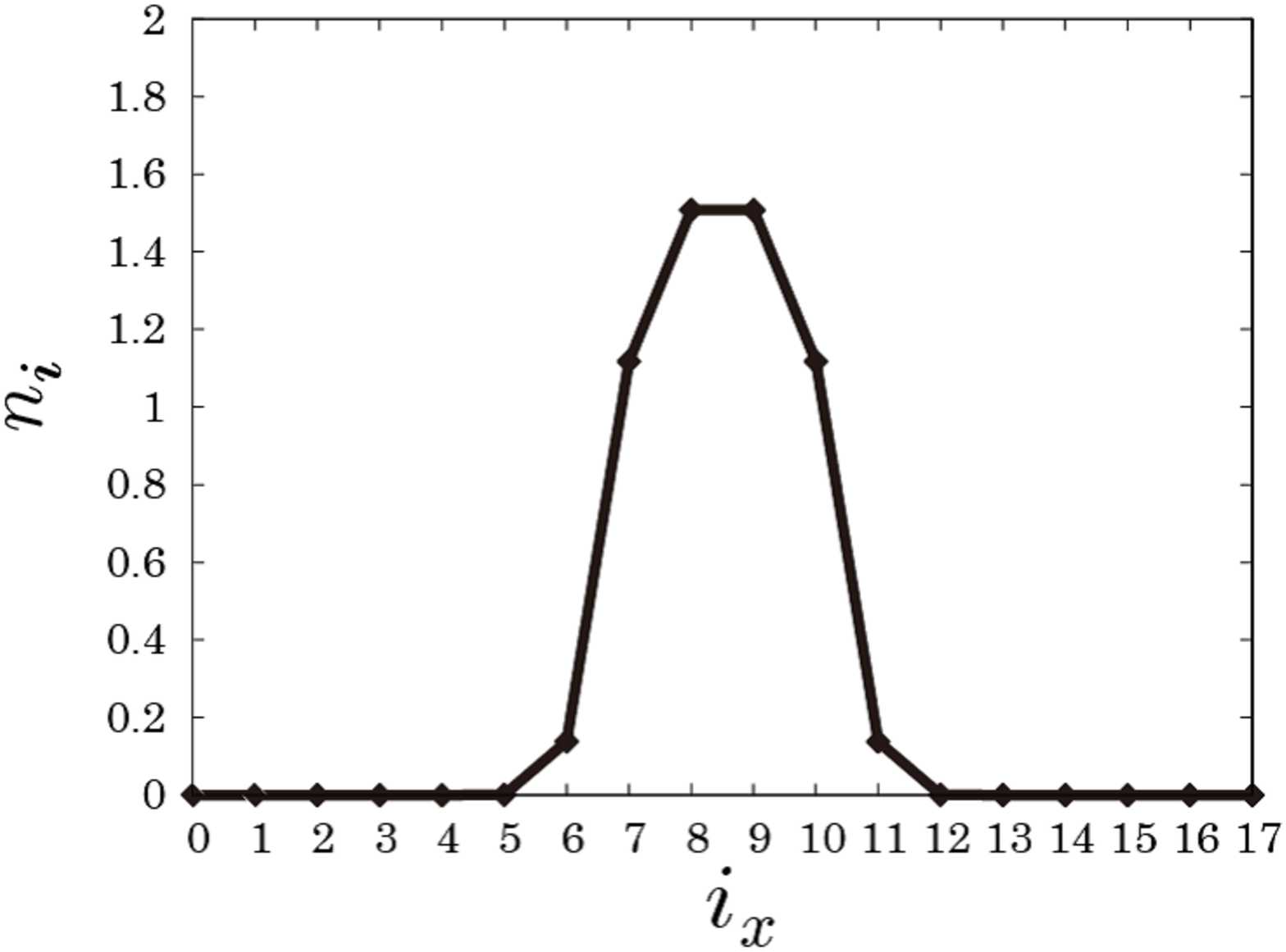}\\
\includegraphics[width=7.2cm]{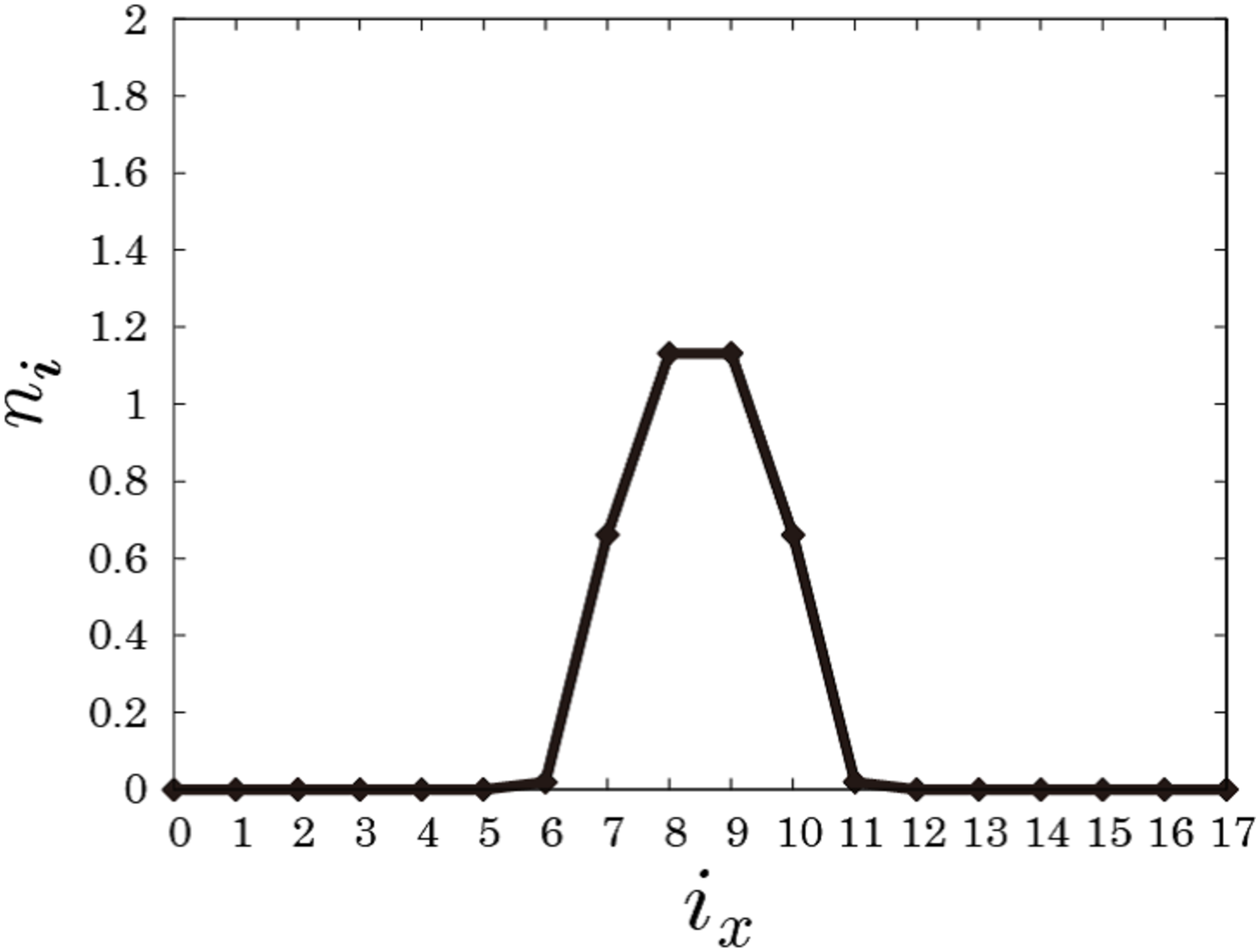}
\caption{
Profiles of initial states: The particle number distributions are shown for $x$-axis at 
$i_y=i_z=9$. The particle number at site $\bi$ is $n_{\bi}=\langle \Psi_G|\hat{n}_{\bi,\uya}+\hat{n}_{\bi,\dya}|\Psi_G \rangle$. Figures (a), (b) and (c) correspond to $N=200, 60, 30$, respectively.}\label{syoki}
\end{figure}

The initial states at $t=0$ are the ground ones under the lattice potential ${\bar V}_0$,
and their profiles are shown for the particle number $N=\sum_{\bi,s}\langle\Psi_G|\hat{a}_{\bi,s}^\dagger\hat{a}_{\bi,s}|\Psi_G\rangle=200, 60, 30$ in Fig. \ref{syoki}. In our simulations below,
we also prepare the initial states for $N=100, 40$.
The responses for the various particle number states are compared in Subsec.~3.3.
  
 Let us consider the variance in the particle distribution $\sigma^2(t)$,
\be
\sigma^2(t)=\frac{1}{N}\sum_{\bi, s}\Big(|\bi-\bi_0|n_{\bi,s}(t)
-\frac{1}{N}\sum_{\bi^\prime,s^\prime}|\bi^\prime-\bi_0|n_{\bi^\prime,s^\prime}(t)\Big)^2
\,.
\ee
 The initial variances are 5.65 for $N=200$ and 2.55 for $N=30$, respectively.
Generally speaking, the increase in the variance indicates that the particles are tended to be
transferred  to the outer sites, while its decrease implies the particle transfer to the inner sites. We will use the variance $\sigma^2(t)$ as a measure of the particle transfer.

\subsection{Responses to the lattice modulation with various frequencies}

\begin{figure}
\includegraphics[width=7.2cm]{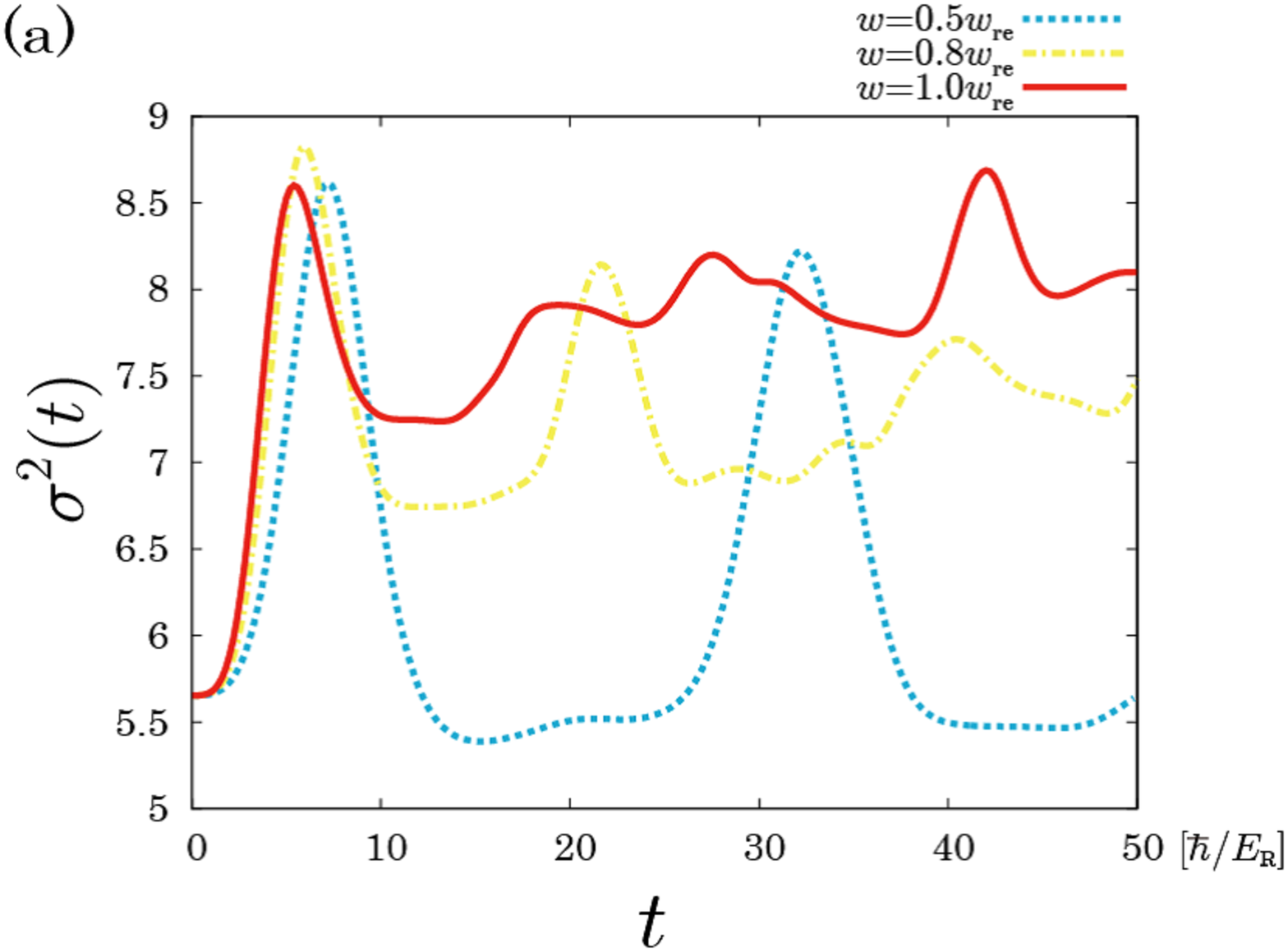}\, \includegraphics[width=7.2cm]{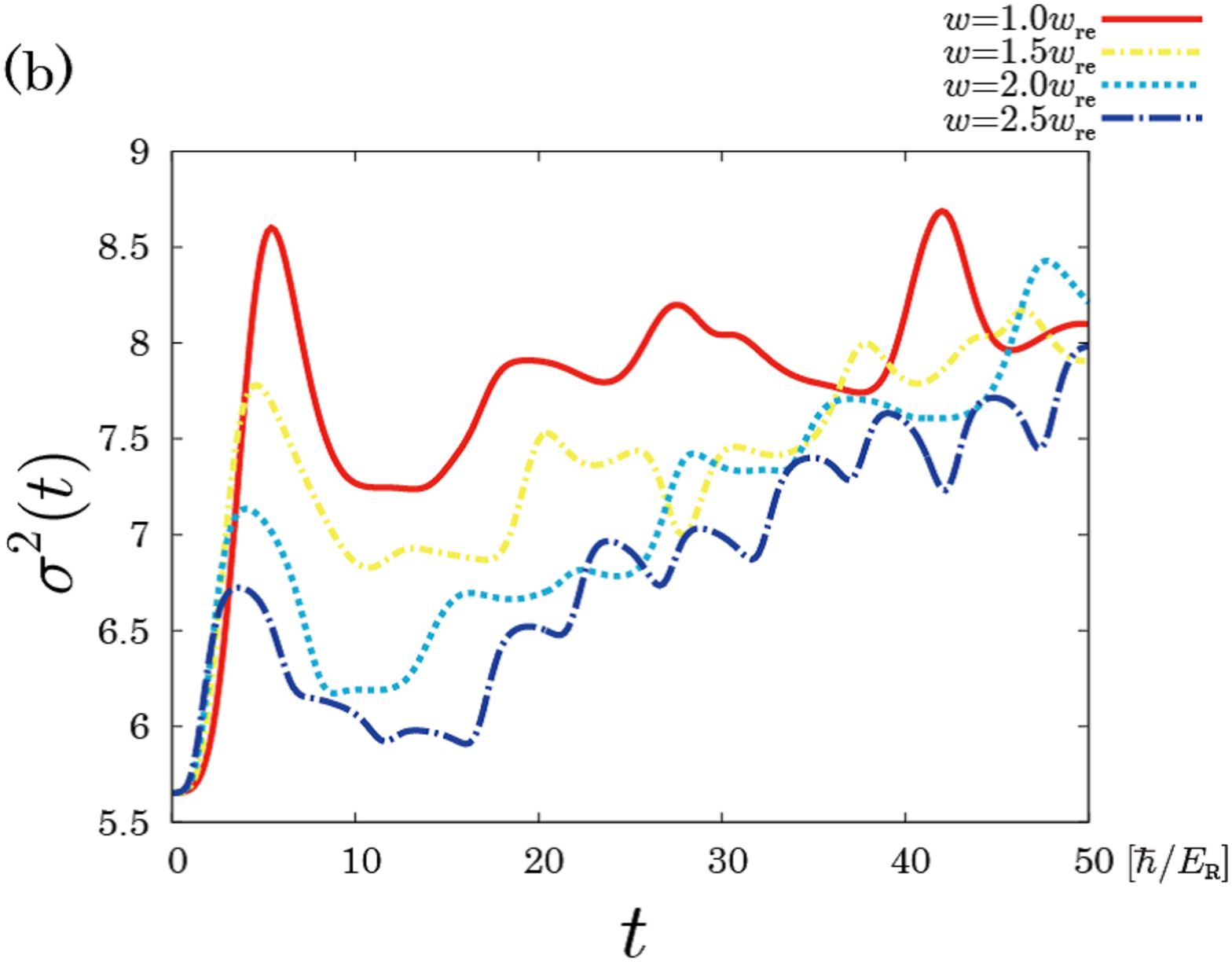}
\caption{Time evolution of the variances in the particle distribution 
for various modulation frequencies with the fixed
modulation amplitude $\delta V/{\bar V_0}=0.5$ and initial phase $\theta=0$ :
 (a) $\omega/\omega_{\rm re}=0.5,\, 0.8,\, 1.0$, 
(b) $\omega/\omega_{\rm re}=1.0,\, 1.5,\, 2.0,\, 2.5$. 
}\label{re-nonre}
\end{figure}
\begin{figure}
\includegraphics[width=7.2cm]{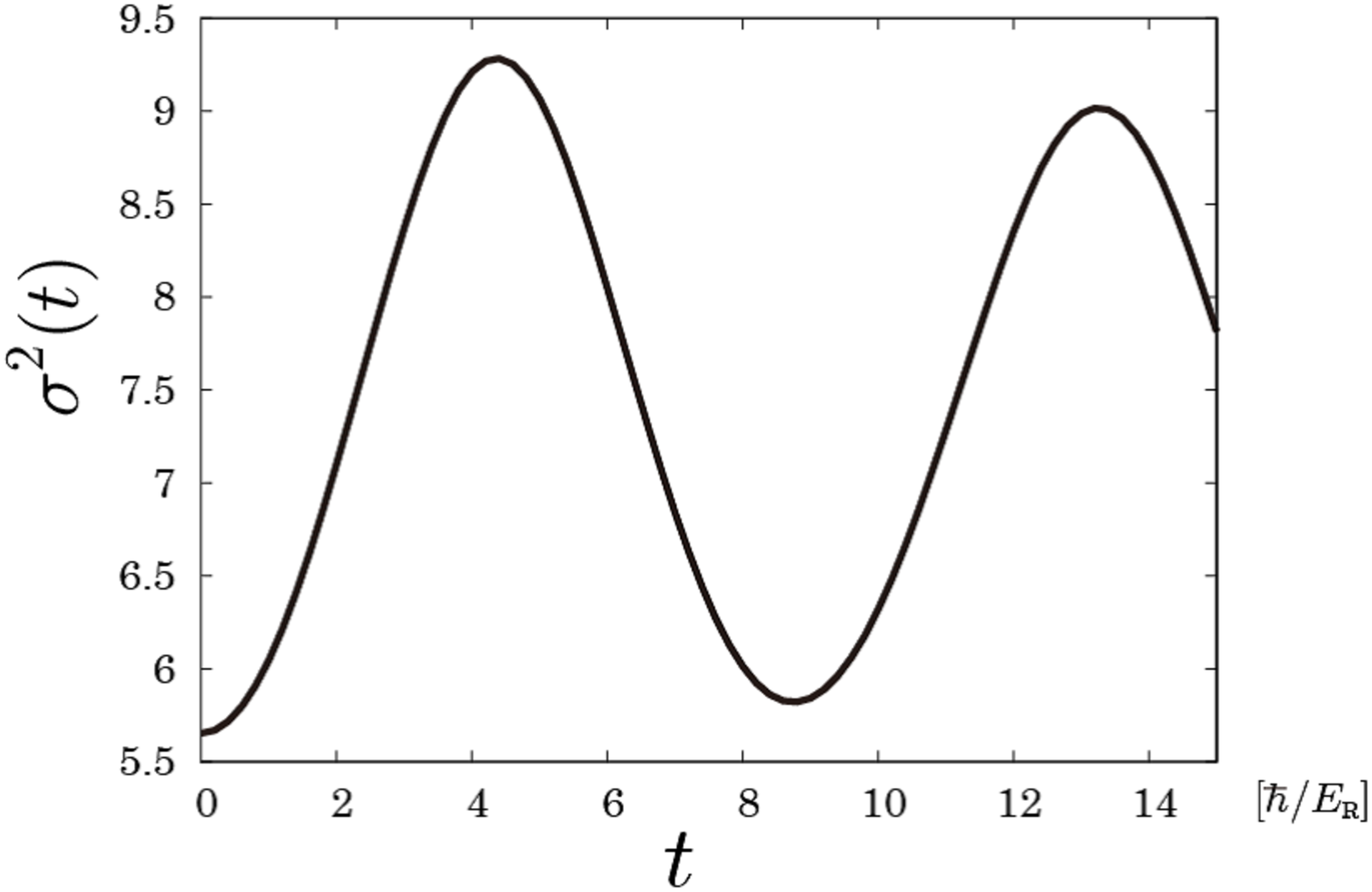}
\caption{Time evolution of the variances in the particle distribution after a sudden change
of the lattice amplitude from ${\bar V}_0=6.5$ to ${\bar V}_0=3.25$ at $t=0$ with $\delta V=0$
and $N=200$.
}\label{int-vib}
\end{figure}

First, we show the time evolution of the variances in the particle distribution 
for various modulation frequencies, fixing the modulation amplitude and the initial phase. 
The initial state is that in 
Fig.~\ref{syoki} (a) ($N=200$). The results are shown for lower frequencies
 in Fig.~\ref{re-nonre} (a)
 and for higher ones in Fig.~\ref{re-nonre} (b), separately.
The fact that the variance $\sigma^2(t)$ becomes large
in general indicates that the particles are transferred to the outer sites,
responding to the lattice modulation. This is the most significant at the resonant
frequency, though it takes place even at non-resonant frequencies.

We see in Fig.~\ref{re-nonre} (a) that the structure of a peak and a plateau is
repeated for the off-resonant frequencies ($\omega= 0.8$ and $0.5 \omega_{\rm re}$).
The time intervals between their first and second peaks
are roughly the respective modulation period $\tau$, which are  $\tau=16.1$ and 
$25.8\hbar/E_{\rm R}$ corresponding $0.8\omega_{\rm re}$ and $0.5\omega_{\rm re}$,
respectively. In order to interpret this, we 
give the temporal behavior of the variance after the lattice amplitude ${\bar V}_0$
is suddenly reduced to a half at $t=0$ (without the lattice modulation),
which is illustrated in Fig.~\ref{int-vib}. There 
the oscillation is understood as follows: Due to the sudden decrease of the lattice
potential height, it is easier for the particles to hop to the neighboring sites, and
at first the particles tend to hop to the outer neighboring sites, since the particle
population is higher at the inner sites, and then go back to the inner ones. It is difficult
to compute the period of this oscillatory behavior, denoted by $T$, analytically,
because many factors take part in it. The numerical calculation in Fig.~\ref{int-vib}
shows $T\simeq 8.8 \hbar/E_{\rm R}$. For $\omega= 0.8$ and $0.5 \omega_{\rm re}$,
the first oscillation peak appears as in Fig.~\ref{int-vib},
but then the lattice potential $V_0(t)$ becomes higher than ${\bar V}_0$ and approaches 
to its maximum value $9.75E_{\rm R}$ for which the lattice system is in the stable insulator.
Therefore, the hopping to the neighboring sites is suppressed extremely,
which explains the plateau. On the other hand, in case of $\omega= 1.0\omega_{\rm re}$,
giving $\tau=12.9\hbar/E_{\rm R}$, $T$ and $\tau$ are of the same order, and the suppression
of the hopping does not last long so that no clear plateau is seen.

In Fig.~\ref{re-nonre} (b), the peak value of the first oscillation is the largest
for the resonant frequency and becomes smaller as the frequency is higher.
The variance continues to increase with the oscillation in all the cases,
implying that the particles mainly moves to the outer sites and scarcely returns
towards the center of the lattice system.  

\begin{figure}
\includegraphics[width=7.2cm]{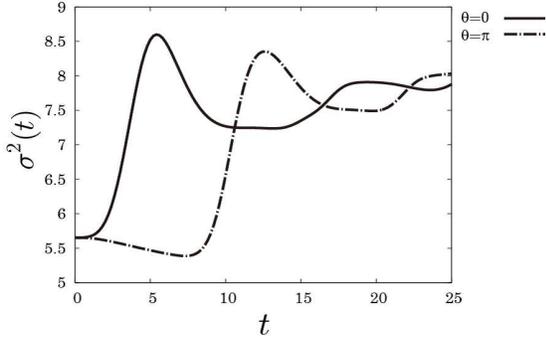}
  \caption{The time evolution of the variances  for the initial phases of the
lattice modulation $\theta=0$ and $\pi$ with $\omega/\omega_{\rm re}= 1.0,\,
\delta V/{\bar V}_0=0.5$ and $N=200$.  
}\label{zure}
\end{figure}

The time evolution of the variance $\sigma^2(t)$ should be
 compared for different $\theta$,
concretely $\theta=0$ and $\pi$, and the result is depicted in Fig.~\ref{zure}.
 As is expected, both behave differently around the initial time
but the difference gradually decreases as $t$ goes. 

In summary of this subsection, the modulation frequency is crucial for the particle
transfer.  The transfer proceeds for the resonant frequency and higher ones, and 
 is the most efficient and rapid for the resonant one. The intrinsic 
oscillation period $T$ after a sudden decrease in the lattice height is an essential
parameter to inhibit the transfer for lower frequencies. The initial phase $\theta$
affects behaviors only for a short interval from the initial time. 

%
%

\subsection{Responses to the lattice modulation with various amplitudes at the resonant frequency} 

\begin{figure}
\includegraphics[width=8cm]{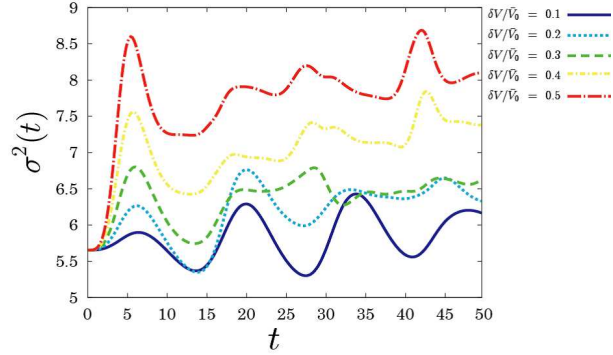}
\caption{Time evolution of the variance in the particle number distribution for various
 lattice amplitude $\delta V/{\bar V}_0 = 0.1,\, 0,2,\, 0.3,\, 0.4$, and $ 0.5$
with $\omega/\omega_{\rm re}= 1.0,\,
\theta=0$ and $N=200$. 
}\label{bunsan}
\end{figure}

 Next, let us see how the response depends on the modulation amplitude $\delta V$.
 The time evolution of the variance in the particle number distribution
 for various values of $\delta V$ at the resonant frequency 
is depicted in Fig.~\ref{bunsan}. It is seen from Fig.~\ref{bunsan}
that the variance follows a simple harmonic-like oscillation when $\delta V$
is small and that for the larger $\delta V$ it
oscillates only for half a period and then increases in a stepwise pattern.

 The particle hopping is suppressed when the lattice amplitude is so large that the stable insulator is formed, but is stimulated when the lattice amplitude is not so large.
 The value of the mean amplitude $\bar{V}_0=6.5E_{\rm R}$ is not large, 
and the particles hop to the outer sites and then to the inner ones periodically
for the lower $\delta V$. For the higher $\delta V$, the particle hopping 
is suppressed as $V_0(t)$ reaches its maximum, though it is 
stimulated around the minimal value of $V_0(t)$.  The stimulation and suppression become more
intensive as $\delta V$ becomes larger. This explains the complicated 
behaviors of the variances for $\delta V/{\bar V_0}= 0.5$ and $0.4$ in Fig.~\ref{bunsan}.
 
\subsection{Dependence of the response on particle number}

\begin{figure}
\includegraphics[width=8cm]{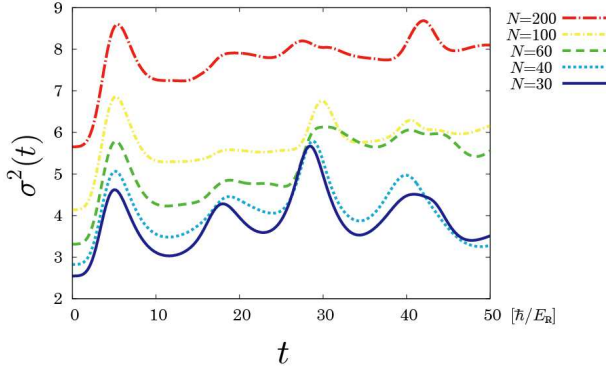}
\caption{Time evolution of the variance in the particle number distribution for various particle number $N=200, 100, 60, 40$, and $30$ with $\omega/\omega_{\rm re}=1.0$, $\theta=0$, and $\delta V/\bar{V}_0=0.5$.
}\label{number-dispersion}
\end{figure}
 
Finally we see how the response depends on the particle number.
 The time evolution of the variance in particle number distribution for various particle 
number $N$ with $\delta V=0.5 {\bar V}_0$ and on the resonant frequency $\omega=\omega_{\rm re}$
is shown in Fig. \ref{number-dispersion}.
 The variance follows a simple harmonic-like oscillation when the particle number is small
 ($N=30$), and oscillates only for half a period and then increases in a stepwise pattern 
for the large particle number such as for $N=200$. 
This is explained as follows: the lattice system  becomes a stable insulator more likely
due to the on-site particle interaction and the Pauli blocking as the particle population
is higher. For the interval in which $V_0(t)$ is large and the stable insulator is formed,
the hopping to the neighboring sites is suppressed and the variance does not change,
meaning the plateau structure. In contrast, the stable insulator is formed hardly for the lower particle population, and the hopping to the neighboring sites is scarcely suppressed for the large 
value of $V_0(t)$, and the variance change in a harmonic-like oscillation.


\section{Summary and Conclusion}

 We consider the ultracold fermionic atom system in a three dimensional optical lattice with 
a confinement harmonic potential,  and analyze it by use of the Hubbard model.
 Our study is focused on the dynamics of the particle transfer. Our numerical calculations
are performed in the Gutzwiller variational approach, because the cost of calculation is
low for higher dimensional and time-dependent systems.
 The lattice modulation frequency is crucial for the particle transfer.

We calculate the variance in the particle distribution, which is a simple quantity
and nevertheless indicates the particle transfer properly.
The results of our numerical calculations are summarized as follows:  The variance changes
most remarkably on the resonant modulation frequency, as quantum transitions are 
enhanced then. On the premise that the mean lattice amplitude ${\bar V}_0$ is close to 
the boundary value at which the stable insulator is formed, the particles are transferred
to the outer sites for the resonant and higher frequencies. For the lower frequencies,
the hopping of the particle to the neighboring sites is suppressed due to the stage
of the stable insulator and the particle transfer is not simulated.  The effect
of the initial phase of the lattice modulation is restrictive.  The modulation amplitude
$\delta V$ is an important parameter. While the variance only oscillates for the lower 
$\delta V$, it continues to increases in a stepwise pattern, leading to the large 
particle transfer. As the total number of atoms $N$ is larger, that is, the particle population
is higher, a stable insulator is formed more easily. This implies that 
the particle transfer is stimulated for the larger $N$, similarly 
as it is so for the larger $\delta V$. 

In conclusion, it is useful to apply the lattice modulation to the fermionic system
in a three dimensional optical lattice with a confinement harmonic potential 
in order to give rise to the particle transfer.  The lattice modulation method is 
quite effective, when the modulation frequency is set to the resonant one, 
and the modulation amplitude and the particle number are large.
The particle transfer can be controlled by the setup conditions on the 
parameters of the lattice modulation, and their proper management makes it possible
to do more complicated experiments of fermionic systems in an optical lattice
under control.

\end{document}